\documentstyle[12pt,epsfig,rotating]{article}
\newcommand\bbb{\ifmmode b\bar{b}\else$b\bar{b}~$\fi}
\newcommand\ttb{\ifmmode t\bar{t}\else$t\bar{t}~$\fi}
\newcommand\Bd{\ifmmode \mathrm B^0_d\else$\mathrm B^0_d~$\fi}
\newcommand\Bs{\ifmmode \mathrm B^0_s\else$\mathrm B^0_s~$\fi}
\newcommand\Dspi{\ifmmode \mathrm D^-_s \pi^+ \else$\mathrm D^-_s \pi^+$\fi}
\newcommand\Dsa{\ifmmode \mathrm D^-_s a_1^+ \else$\mathrm D^-_s a_1^+$\fi}
\newcommand\Ds{\ifmmode \mathrm D^-_s \else$\mathrm D^-_s $\fi}
\newcommand\Ks{\ifmmode \mathrm K^0_s\else$\mathrm K^0_s~$\fi}
\newcommand\xs{\ifmmode x_{\mathrm s}\else$x_{\mathrm s}~$\fi}
\newcommand\jpsi{\ifmmode J/\psi\else$J/\psi~$\fi}
\newcommand\lu{\ifmmode {\mathrm cm^{-2}s^{-1}}\else$
{\mathrm cm^{-2}s^{-1}}~$\fi}
\newcommand\lmax{\ifmmode 1.7 \cdot 10^{34}\else$1.7 \cdot 10^{34}~$\fi}
\newcommand\lhigh{\ifmmode 10^{34}\else$10^{34}~$\fi}
\newcommand\lmed{\ifmmode 5 \cdot 10^{33}\else$5 \cdot 10^{33}~$\fi}
\newcommand\llow{\ifmmode 10^{33}\else$10^{33}~$\fi}
\newcommand\ipb{\ifmmode {\mathrm pb}^{-1}\else${\mathrm pb}^{-1}$\fi}
\newcommand\pT{\ifmmode p_{\mathrm T}\else$p_{\mathrm T} $\fi}
\newcommand\ra{\ifmmode \rightarrow\else$\rightarrow$\fi}
\newcommand\rfi{\ifmmode R\varphi \else$R\varphi~$\fi}
\newcommand\um{\ifmmode \mu {\mathrm m} \else$\mu$m \fi}
\newcommand\st{\ifmmode \sigma_t \else$\sigma_t$\fi}
\newcommand\xsm{\ifmmode x_s^{max} \else$x_s^{max}$\fi}
\newcommand\lint{\ifmmode \int L dt \else$\int L dt$ \fi}
\newcommand\lintu{\ifmmode \mathrm 10^4\ pb^{-1}  \else$\mathrm 10^4\ pb^{-1}$ \fi}
\newcommand\bd{\begin{description}}
\newcommand\ed{\end{description}}

\begin{document}
\hfill
\parbox{5.5cm}{{\large HU-SEFT R 1996-21} \\
              {\large October~31,~1996}} \\

\vskip 0.7cm
\begin{center}
{\huge ATLAS sensitivity range} \\
{\huge for the $x_s$ measurement}
\vskip 0.5cm
{\large  S.~Gadomski}
\vskip 0.2cm
{\large \it INP Cracow, Poland}
\vskip 0.3cm
{\large  P.~Eerola}
\vskip 0.2cm
{\large \it CERN, Geneva, Switzerland}
\vskip 0.3cm
{\large  A.~V.~Bannikov}
\vskip 0.2cm
{\large \it JINR Dubna, Russia}
\end{center}
\vskip 0.7cm

\abstract{Previous results for the prospects of 
$\Bs$ mixing measurement in the ATLAS experiment at LHC are updated.
The improved analysis method of the studied decay channels $\Bs \ra \Dspi$
and $\Bs \ra \Dsa$,
combined with most recent values for the 
branching ratios and the \Bs lifetime,
leads to the new \mbox{ATLAS} sensitivity 
range for the $x_s$ measurement: $x_s^{max} = 42$. An extensive
study is done in order to estimate how $x_s^{max}$ is influenced
by the B-decay proper-time resolution of the vertex detector, 
as well as by the number of events and by the signal-to-background 
ratio.}


\section{
\boldmath 
Measurement of $x_s$
\unboldmath
\label{explim}}

CP-violation in B-meson decays can be characterized by the unitarity triangle,
composed of Cabibbo-Kobayashi-Maskawa matrix elements by requiring 
that the matrix is unitary. 
In addition to the angles of the unitarity triangle, measurements of the
lengths of the sides provide us with complementary information about the
triangle. The length of the side $|V_{td}|$ is least well-known of the sides.
The mixing parameter $x_{\rm d}$ is proportional to $|V_{td}|^2$,
but inferring the value of $|V_{td}|^2$ from $x_{\rm d}$
is hampered by hadronic uncertainties.
To large extent, these uncertainties
cancel, when considering the ratio of the mixing parameters, 
$x_{\rm s}/x_{\rm d} \propto |V_{ts}/V_{td}|^2$.

The mixing parameter $x_{\rm s}$ has not been measured yet. The Standard Model
predicts it to be in the range 10-30, and the present lower limit from LEP
is \mbox{$x_{\rm s} > 11.1$~\cite{zeitnitz}}.

\begin{figure}
\begin{center}
\mbox{\epsfig{file=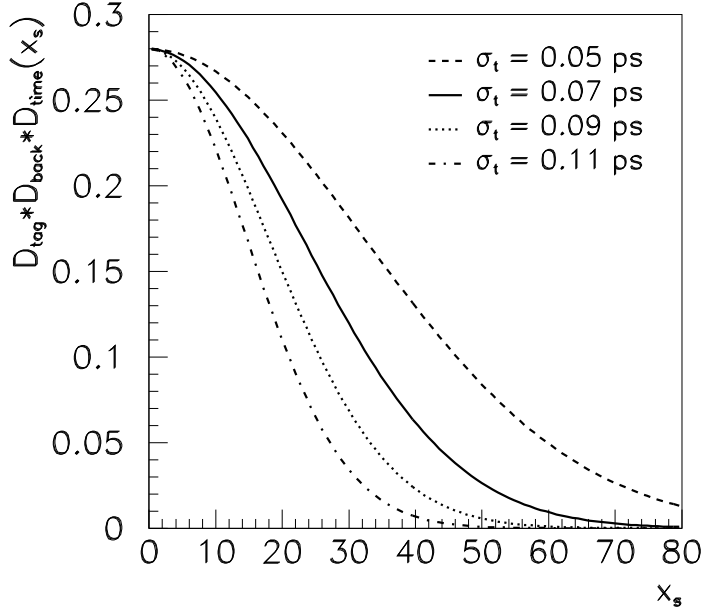,width=10.0cm}}
\caption{Expected amplitude of the asymmetry 
$D = D_{tag} D_{back} D_{time}(x_s)$ 
as a~function of $x_s$ for various proper
time resolutions $\sigma_t$. 
\label{eampl}}
\end{center}
\end{figure}
Principles of the $x_s$ measurement are described in previous notes
\cite{chan1,chan2}. A time-dependent asymmetry $A(t)$, observed in 
the experiment, is defined as:
\begin{equation}
A(t)=\frac{dn(++)/dt-dn(+-)/dt}{dn(++)/dt+dn(+-)/dt}=D \cos(x_s t/\tau_{B_s})
\label{fasy}
\end{equation}
where $(++)$ refers to B-decays in which the flavour of the B appears
to be the same in the production and decay, $(+-)$ refers to cases where 
an oscillation of the B into its anti-particle is observed, and $t$ is 
the proper time. $D$ is a~product of all the dilution factors:
\begin{eqnarray}
D & = & D_{tag} D_{back} D_{time} \label{dtotal} \\
\nonumber \\
D_{tag} & = & 1 - 2W\\
D_{back} & = & \frac{N_S}{N_S + N_B} \\
D_{time} & = & \exp{(-\frac{1}{2}(\frac{\sigma_t x_s}{\tau})^2)} 
\label{dtime} 
\end{eqnarray}
where $W$ is the wrong tag fraction, $N_S$ is the number of signal events,
$N_B$ is the number of background events\footnote{It is assumed 
for the formula \ref{fasy} that 
background $dN_B/dt \sim exp(-t/\tau)$, with no asymmetry.},    
$\sigma_t$ is the proper time resolution and $\tau$ is the 
\Bs lifetime. Formula \ref{dtime} can be obtained using Fourier transforms 
\cite{hgm}. 

In case of \mbox{ATLAS} \- $W = 0.22$ \cite{TP} and $N_B = N_S$ is assumed
(section \ref{params}).
The lifetime of the \Bs is assumed to be 
$\tau = 1.61$~ps after \cite{PDG96}.
Figure \ref{eampl} shows the total dilution factor $D$ estimated
with the above assumptions for different values of the proper-time
resolution ($\sigma_t = 0.05$, 0.07, 0.09 and 0.11 ps), as a~function   
of $x_s$. The amplitude of the cosine wave which can be observed by
the experiment is reduced with increasing $x_s$ because of the finite
$\sigma_t$, as given by the formula \ref{dtime}. One can expect that
for any given statistical sensitivity of the experiment, defined by
the available signal statistics and by the background, there is a~minimal
value of the asymmetry amplitude $D$ which still enables a~reliable 
measurement of $x_s$. Therefore, from equation \ref{dtime}, one expects 
\begin{equation}
x_s^{max} \sim \frac{1}{\sigma_t} \label{basic}
\end{equation}
for any objective definition of $x_s^{max}$.   

\section{Definition of the \mbox{ATLAS} range \label{sectrange}}

\subsection{Input parameters} 
\label{params}

In order to estimate the sensitivity of the experiment for the $x_s$
measurement, we use as input parameters the number of signal events $N_S$,
the number of background events $N_B$, and the resolution in the 
proper-time of the \Bs decay, \st. 

Two of the \Bs decay channels which can be used for the $x_s$ measurement
have been analyzed so far: $\Bs \ra \Dspi$ and $\Bs \ra \Dsa$. 
The original work is described in \cite{chan1} and in \cite{chan2},
respectively. The results have been updated, for the Technical Proposal
and on a~few other occasions, because of minor changes of the cuts,
updated branching ratios or trigger efficiencies, in particular 
the second level trigger sensitive to $\mathrm D_s^\pm$. 
It is therefore useful to summarize the event counting as it was done 
for this work. The calculation is explained in tables \ref{events1}
and~\ref{events2}.

\begin{table}[tb]
\caption{Number of signal events from $\Bs \ra \Dspi$ channel
expected in \mbox{ATLAS} after 1~year ($10^7$ s) of operation at 
\llow \lu
\label{events1}}
\begin{center}
{\small
\begin{tabular}{|l|c|l|}\hline
Parameter & Value  & Comment \\ \hline
${\cal L}$ [cm$^{-2}$s$^{-1}$] & $10^{33}$ & \\
$t$ [s] & $10^7$ & \\
$\sigma({\mathrm {b\bar{b}}} \rightarrow \mu$X) [$\mu$b] & 2.3 &
$p_{\mathrm T}^\mu >$ 6 GeV/$c$\\
& & $|\eta^\mu| <$ 2.2 \\ \hline
$N({\mathrm {b\bar{b}}} \rightarrow \mu$X) & $2.3 \cdot 10^{10}$ & \\ \hline
Br($\mathrm{b \rightarrow B_s^0}$) & 0.112 & \\
Br($\mathrm{B_s^0 \rightarrow D_s^-\pi^+}$) & 0.003 & \\
Br($\mathrm{D_s^- \rightarrow \phi^0 \pi^-}$) & 0.036 & \\
Br($\mathrm{\phi^0 \rightarrow K^+ K^-}$) & 0.491 & \\ \hline
$N(\mathrm{K^+K^-} \pi^+\pi^-)$ & 136,600 & \\ \hline
Acceptance of the cuts: & & \\
$p_{\mathrm T} > 1$~GeV/$c$ & & \\
$|\eta| < 2.0$ & & \\
$\Delta \varphi_{\mathrm{KK}} < 10^o$ & & \\
$\Delta \Theta_{\mathrm{KK}} < 10^o$ & & \\
$|M_{\mathrm{KK}} - M_{\phi^0}| < 10$~MeV/$c^2$ & & \\
$|M_{\mathrm{KK\pi}}- M_{\mathrm {D_s^-}}| <
15$~MeV/$c^2$ & & \\ 
$\mathrm {D_s^-}$ vertex fit $\chi^2 < 10.0$ & & \\
$\mathrm {B_s^0}$ vertex fit $\chi^2 < 5.0$ & & \\
$\mathrm {B_s^0}$ proper decay time $>$ +0.4 ps & & \\
$\mathrm {B_s^0}$ impact parameter $< 55 \mu$m & & \\
$\mathrm {B_s^0}$  $ p_{\mathrm T} > 10.0$~GeV/$c$ & 8\% & \\
\hline
$N(\mathrm{K^+K^-}\pi^+\pi^-)$ after cuts & 10,900 & \\ \hline
Trigger efficiency & 0.54 & \\
Lepton identification & 0.8 & \\
Track efficiency & $(0.95)^4$ & \\ 
Mass cut $\pm 2 \sigma$ & 0.95 & \\ \hline
$N(\mathrm {K^+K^-}\pi^+\pi^-)$ reconstructed & 3,640 & \\ \hline
\end{tabular}
}
\end{center}
\end{table}

\begin{table} [tb]
\caption{Number of signal events from $\Bs \ra \Dsa$ channel
expected in \mbox{ATLAS} after 1~year ($10^7$ s) of operation at 
\llow \lu
\label{events2}}
\begin{center}
{\small
\begin{tabular}{|l|c|l|}\hline
Parameter & Value  & Comment \\ \hline
${\cal L}$ [cm$^{-2}$s$^{-1}$] & $10^{33}$ & \\
$t$ [s] & $10^7$ & \\
$\sigma({\mathrm {b\bar{b}}} \rightarrow \mu$X) [$\mu$b] & 2.3 &
$p_{\mathrm T}^\mu >$ 6 GeV/$c$\\
& & $|\eta^\mu| <$ 2.2 \\ \hline
$N({\mathrm {b\bar{b}}} \rightarrow \mu$X) & $2.3 \cdot 10^{10}$ & \\ \hline
Br($\mathrm{b \rightarrow B_s^0}$) & 0.112 & \\
Br($\mathrm{B_s^0 \rightarrow D_s^-a^+_1}$) & 0.006 & \\
Br($\mathrm{D_s^- \rightarrow \phi^0 \pi^-}$) & 0.036 & \\
Br($\mathrm{\phi^0 \rightarrow K^+ K^-}$) & 0.491 & \\ 
Br($\mathrm{a^+_1 \rightarrow \rho^0 \pi^+}$) & $\sim 0.5$ & \\ 
Br($\mathrm{\rho^0 \rightarrow \pi^+ \pi^-}$) & $\sim 1$ & \\ 
\hline
$N(\mathrm{K^+K^-} \pi^+\pi^- \pi^+\pi^-)$ & 136,600 & \\ 
\hline
Acceptance of the cuts: & & \\
$p_{\mathrm T} > 1$~GeV/$c$ & & \\
$|\eta| < 2.5$ & & \\
\hline
$N(\mathrm{K^+K^-} \pi^+\pi^- \pi^+\pi^-)$ & 9,015 & 6.6\% \\
\hline
$\Delta \varphi_{\mathrm{KK}} < 10^o$ & & \\
$\Delta \Theta_{\mathrm{KK}} < 10^o$ & & \\
$|M_{\mathrm{KK}} - M_{\phi^0}| < 20$~MeV/$c^2$ & & \\
$|M_{\mathrm{KK\pi}}- M_{\mathrm {D_s^-}}| <
15$~MeV/$c^2$ & & \\ 
$\Delta \varphi_{\pi\pi} < 35^o$ & & \\
$\Delta \Theta_{\pi\pi} < 15^o$ & & \\
$|M_{\pi\pi} - M_{\rho^0}| < 192$~MeV/$c^2$ & & \\
$|M_{\mathrm{\pi\pi\pi}}- M_{\mathrm {a^+_1}}| < 300$~MeV/$c^2$ & & \\ 
\hline
$N(\mathrm{K^+K^-} \pi^+\pi^- \pi^+\pi^-)$ & 6,830 & 5.0\% \\
\hline
$\mathrm {D_s^-}$ vertex fit $\chi^2 < 12.0$ & & \\
$\mathrm {a^+_1}$ vertex fit $\chi^2 < 12.0$ & & \\
$\mathrm {B_s^0}$ proper decay time $>$ +0.4 ps & & \\
$\mathrm {B_s^0}$ impact parameter $< 55 \mu$m & & \\
$\mathrm {B_s^0}$  $ p_{\mathrm T} > 10.0$~GeV/$c$ & 8\% & \\
\hline
$N(\mathrm{K^+K^-}\pi^+\pi^-\pi^+\pi^-)$ after cuts & 4,100 & 3.0\% \\ \hline
Trigger efficiency & 0.54 & \\
Lepton identification & 0.8 & \\
Track efficiency & $(0.95)^6$ & \\ 
Mass cut $\pm 2 \sigma$ & 0.95 & \\ \hline
$N(\mathrm {K^+K^-}\pi^+\pi^-\pi^+\pi^-)$ reconstructed & 1,240 & \\ \hline
\end{tabular}
}
\end{center}
\end{table}

The number of background events for $\Bs \ra \Dspi$ was also
estimated in \cite{chan1}. First some potentially dangerous
exclusive decays (${\mathrm {B^0_d \rightarrow D^-_s} \pi^+}$,
${\mathrm {B^0_d \rightarrow D^- \pi^+}}$
and ${\mathrm {\Lambda_b \rightarrow \Lambda^+_c \pi^-}}$ followed by
${\mathrm {\Lambda^+_c \rightarrow p K^- \pi^+ }}$) were checked and 
found not to contribute significantly to the background. The combinatorial 
background can not be precisely estimated because of the very 
large Monte Carlo statistics required. 
With the set of cuts shown in table \ref{events1}, no candidate
event was found in the mass window $5 < M_{KK\pi\pi} < 6$~GeV
within 370,000 inclusive $\mu$X events. That gives an upper limit
on the level of $N_B \leq \sim 0.6 \cdot N_S$ within the final mass cut
of  $\pm 2 \sigma$. 

A similar procedure was carried out in \cite{chan2} for the
channel $\Bs \ra \Dsa$. The channels 
$\mathrm B_d^0 \ra D^- a_1^+$,
$\mathrm \Lambda_b \ra  \Lambda^+_c \pi^-$ followed by
$\mathrm \Lambda^+_c \ra p K^- \pi^+ \pi^+ \pi^-$ and
$\mathrm B_d^0 \ra D_s^- a_1^+$ were found not to contribute
significantly, as compared to the limit on combinatorial background,
which was $N_B \leq \sim N_S$. 
The latter was based on the Monte-Carlo sample of 300,000 inclusive 
$\mu$X events, where no event has passed cuts from table \ref{events2}
in the mass window
$M_{\Bs}-150\ {\mathrm MeV/c^2} < M_{\mathrm KK\pi\pi\pi\pi} <
 M_{\Bs}+150\ {\mathrm MeV/c^2}$.

Adding the contribution from the two channels, the total number of signal
events is $N_S = 4480$ for one year of low luminosity data taking 
($\lint = \lintu$). A~conservative value of $N_B = N_S$ was used
in the following. A~dedicated study of the impact of background
on $\xsm$ was also done, and is described in section \ref{depback}. 

The proper-time resolution $\st$ was calculated in \cite{chan1} using
full GEANT simulation of the \mbox{ATLAS} tracker equipped with the strip
'B-physics' layer. The result was $\st = 0.069$~ps. A slightly better
value $\st = 0.064$~ps was obtained with track level simulation
for the second decay channel in \cite{chan2}. 
There are many other track-level results, which were produced later, and
which are in a~good agreement with the latter value. 

Replacing 
the strip 'B-physics' layer at $R = 6$~cm with a~third pixel layer
at $R = 4$~cm gives slightly worse performance, $\st = 0.068$
to 0.094~ps, depending on the resolution assumed for pixels 
\cite{PEJune}.
Removal of the 'B-physics' layer, with the first tracking layer
made of pixels at $R = 11$~cm, results in $\st = 0.146$ to 0.154~ps.

A value of $\st = 0.07$~ps, well justified by the result from \cite{chan1}, 
which is the only one so far using full simulation, is assumed in this 
section. 
In section \ref{depend} the work is repeated for values of $\st$ ranging 
from 0.05 to 0.011~ps. 

A parameter of importance which was also updated for this work is the 
$\Bs$ lifetime. An up to date value $\tau = 1.61$~ps used here is higher
than in case of previous presentations ($\tau = 1.54$~ps, \cite{forty}), 
adding to an increase of the predictions for $\xsm$.  
 
\begin{figure}
\begin{center}
\mbox{\epsfig{file=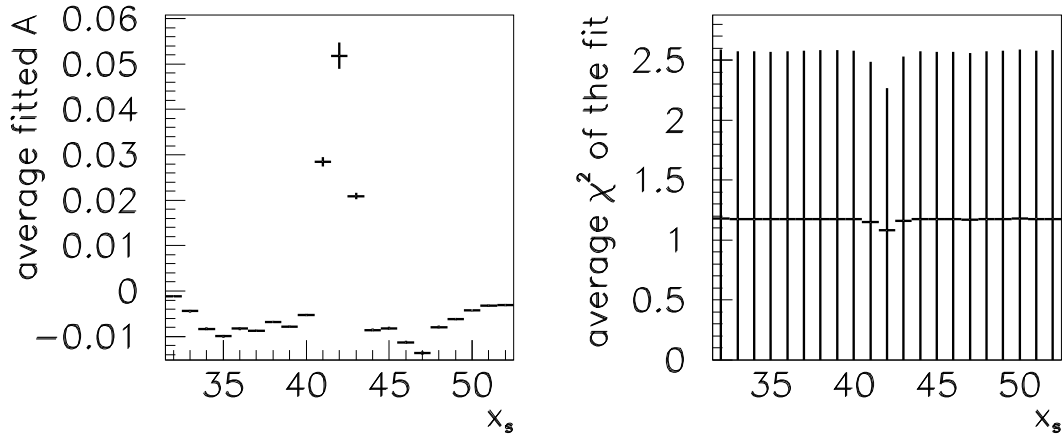,width=15.0cm}}
\caption{An average fitted amplitude $A(x_s)$ (left) and an average $\chi^2$ 
of the fit (right), both as a~function of $x_s$, from 1000 experiments.
The error bars show the RMS.
\label{ampchi3}}
\end{center}
\end{figure}
 
\subsection{Procedure of defining the range}

The number of signal events $N_S$, the number of background
events $N_B$ and the proper-time resolution $\st$ are used
as parameters in a~Monte-Carlo program, which generates
"experimental" distribution of the "measured" asymmetry 
$A(t)$. The distribution then has $D_{back}$, $D_{time}$ and 
Poisson fluctuations introduced with Monte-Carlo methods. 
Dilution from wrong tags $D_{tag}$ is also introduced, with 
the wrong tag fraction $W = 0.22$ \cite{TP}. Finally, 
the program generates $A(t)$ distributions for $t > 0.4$~ps because
of the analysis cut listed in tables \ref{events1} and \ref{events2}.   

Resulting 'experimental' $A(t)$ distribution is analyzed with a~method
called the amplitude fit, recommended in \cite{hgm}. $A(t)$ distribution 
is fitted with the function\\ 
$A_{fit} \cos(x_s t/\tau)$, 
where $x_s$ is a~constant value and $A_{fit}$ is the only 
free parameter. The fit is repeated for different values of $x_s$, giving 
an $A_{fit}(x_s)$ distribution. 

The amplitude fit method can be described as a~variant of the Fourier
transformation, a~'cosine transformation'. It has the power of Fourier 
analysis for the periodical signals, but one introduces known information
about the phase in addition - we are looking explicitly for a~cosine 
wave in the $A(t)$ distribution. 

Figure \ref{ampchi3} shows the properties of the transformation.    
An average fitted amplitude is shown as a~function of $x_s$  
after 1000 experiments which had the true $x_s = 42$. An average $\chi^2$ 
value of the fit is  also shown as a~function of $x_s$ and the error bars show 
the RMS calculated from 1000 'experiments'.
It can be observed,  that $A_{fit}(x_s)$ is more useful to find the true 
value of the oscillation frequency.

The value of $x_s$ giving the highest $A_{fit}$ is the $x_s$ measurement
of the experiment. The peak in the fitted amplitude has its natural
width, which can be seen in Fig.~\ref{ampchi3} (more explanation
can be found in \cite{hgm}). An experiment is called 'successful' here 
when the measured $x_s$ value is within that width from the 'true' value 
defined in the Monte-Carlo. 

With increasing 'true $x_s$' the amplitude of the oscillation seen
in the $A(t)$ distribution decreases, as shown in figure \ref{eampl}. 
When a~limit is passed the amplitude fit no longer enables to distinguish
the right oscillation frequency form the noise generated by the statistical
fluctuations in bins. 

\begin{figure}
\begin{center}
\mbox{\epsfig{file=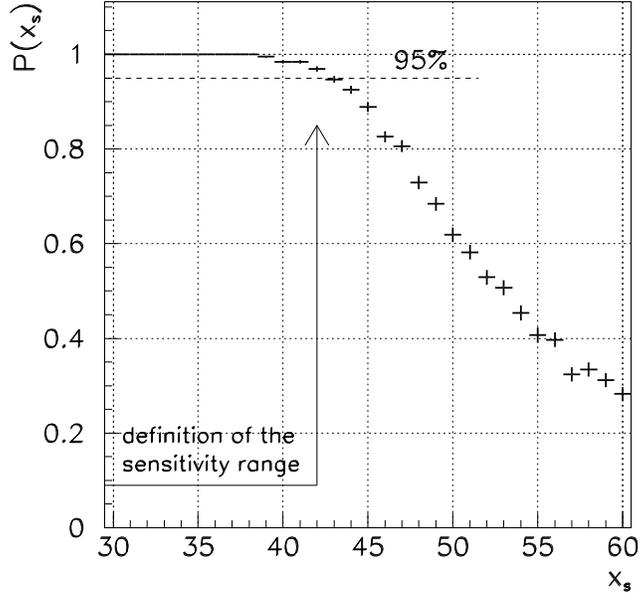,width=10.0cm}}
\caption{Probability of a~successful measurement, estimated using 
1000 "experiments" with nominal \mbox{ATLAS} parameters: $\sigma_t = 0.07$~ps
and $\lint = \lintu$. 
\label{prob}}
\end{center}
\end{figure}

The probability of the experiment success calculated from 1000 simulated
'experiments' is shown in figure \ref{prob} as a~function of $x_s$. 
The limit $x_s^{max}$ is defined as the highest value of $x_s$ 
for which such probability of an experimental success is above 95\%.  

\begin{figure}
\begin{center}
\mbox{\epsfig{file=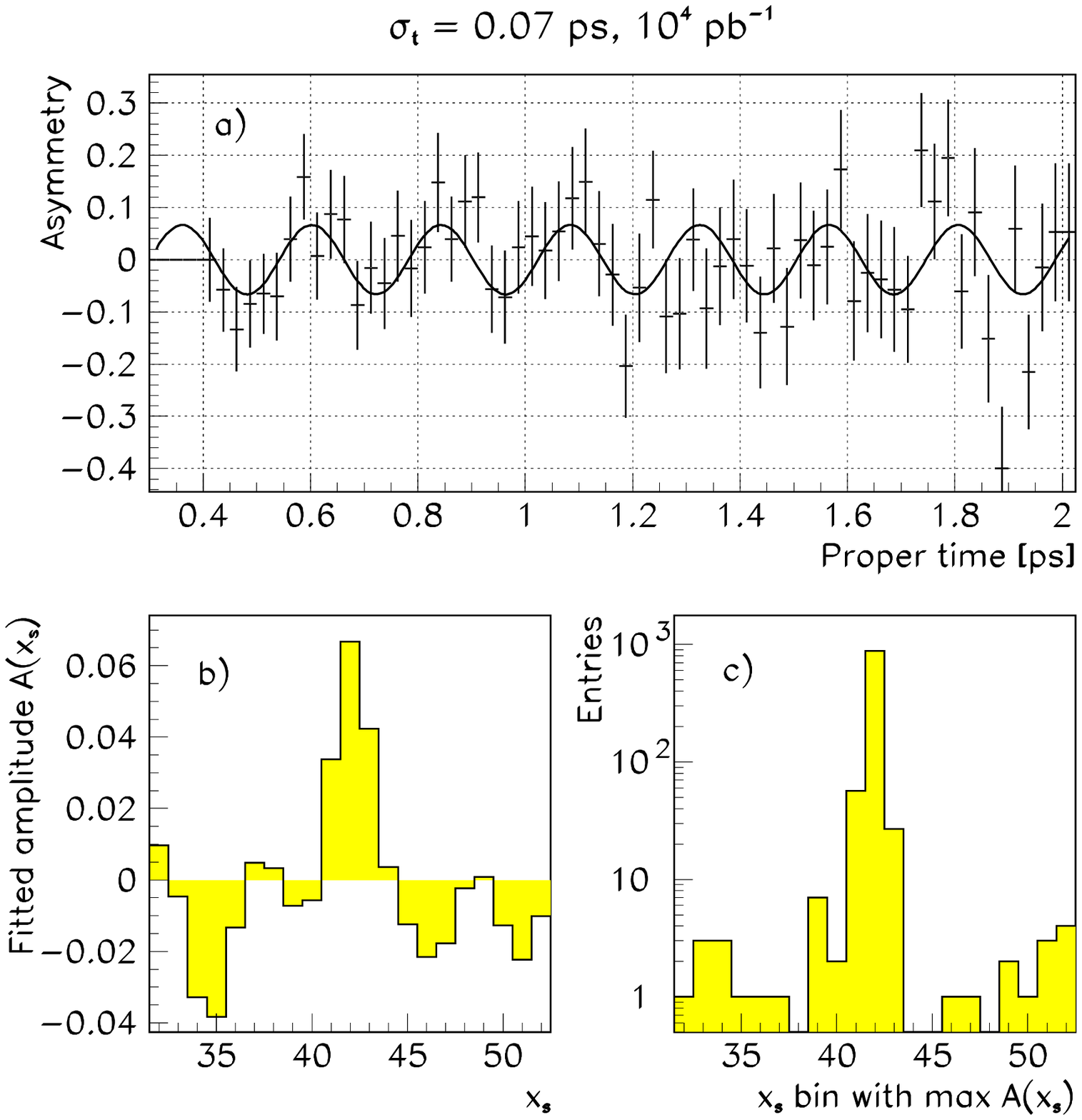,width=17.0cm}}
\caption{$x_s = 42$, "nominal" parameters: $\sigma_t = 0.07$~ps 
and $\lint = \lintu$; (a) time-dependent asymmetry distribution 
(single experiment), (b) amplitude $A(x_s)$ fitted to the distribution
shown in (a), (c) distribution of $x_s$ giving the highest value
of $A(x_s)$ when the experiment was repeated 1000 times. 
\label{amf07y1n}}
\end{center}
\end{figure}

For \mbox{ATLAS}, with the parameters listed in section \ref{params} the limit 
turns out to be $x_s^{max}=42$. A time-dependent asymmetry distribution, 
its transformation with the amplitude fit procedure and the results of
1000 simulated experiments are shown in Fig.\ref{amf07y1n}.

\begin{figure}
\begin{center}
\mbox{\epsfig{file=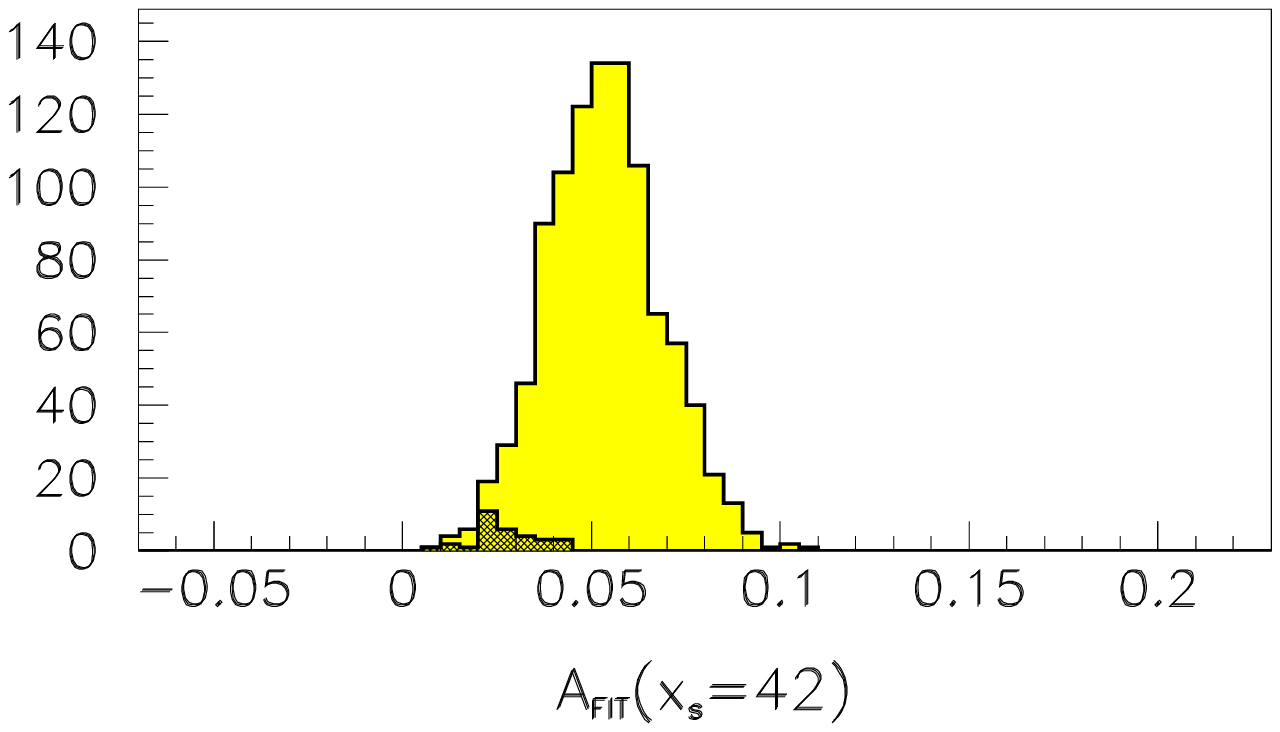,width=12.0cm}}
\caption{Distribution of the amplitudes fitted for $x_s = 42$
for 1000 'experiments' (generated with the 'true' $x_s = 42$).
Dark histogram shows amplitudes for 31 'experiments' which 'fail'.   
\label{ampd}}
\end{center}
\end{figure}

For the correct frequency an average amplitude should be equal to 
the product of all dilution factors $D$ (see formula \ref{fasy}). 
Figure \ref{ampd} shows the distribution of the fitted amplitudes.
An average value $(53.15 \pm 0.49) \cdot 10^{-3}$ 
is in agreement with the prediction from formulas \ref{dtotal} to 
\ref{dtime} equal $D = 52.85 \cdot 10^{-3}$.

\section{\boldmath Dependence of the range on $\sigma_t$ and on 
integrated luminosity \label{depend}}

In order to evaluate the importance of the proper time resolution
$\st$ and of the integrated luminosity, the analysis was repeated
for different values of these parameters. All combinations given
by $\st$ values of 0.05, 0.07, 0.09 and 0.11~ps, together with
\lint values of 0.5, 1.0, 2.0, 5.0 and $10.0 \times \lintu$ were
tried.

\begin{figure}
\begin{center}
\mbox{\epsfig{file=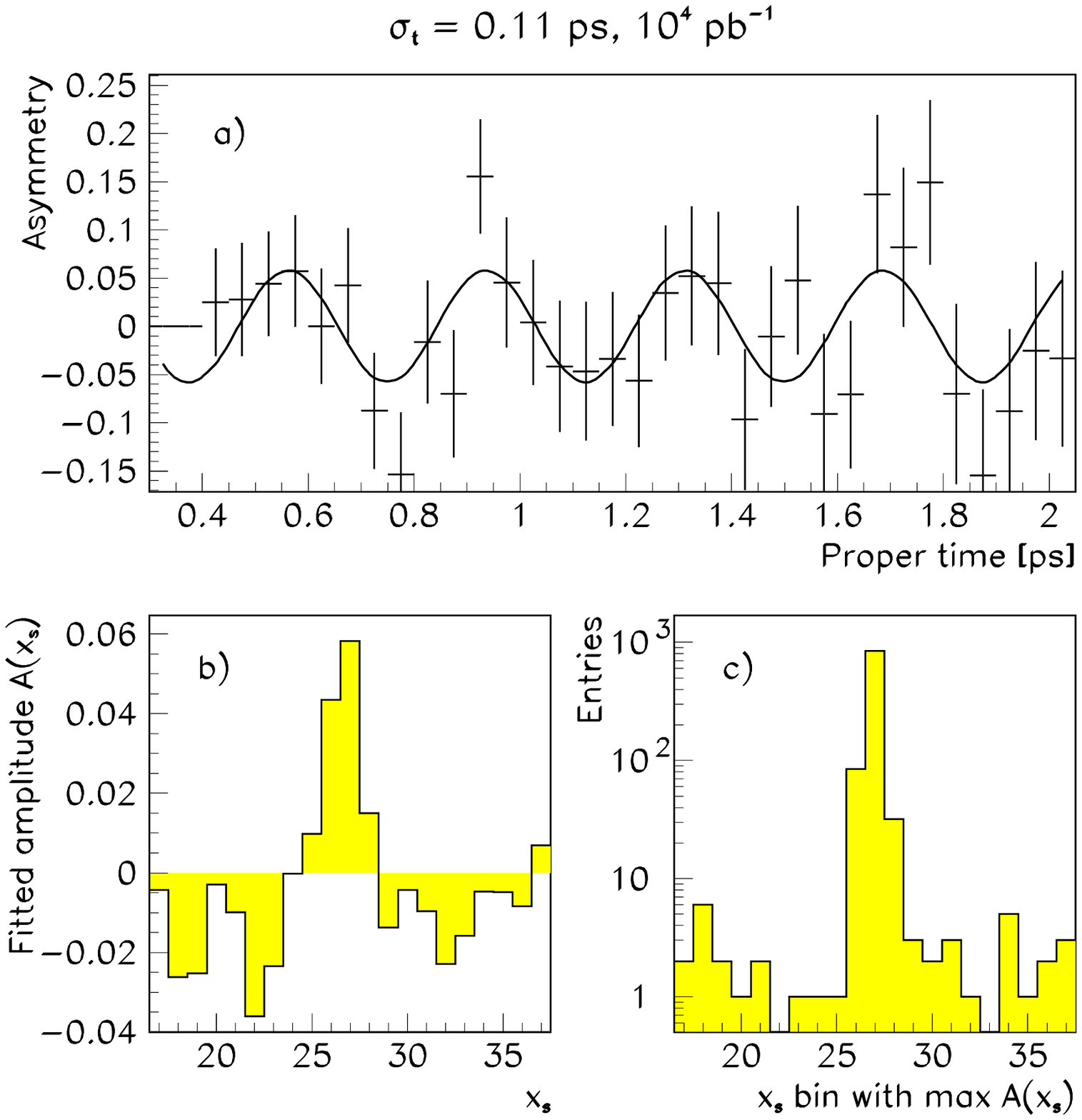,width=17.0cm}}
\caption{$x_s = 27$, $\sigma_t = 0.11$~ps 
and $\lint = \lintu$; (a) time-dependent asymmetry distribution 
(single experiment), (b) amplitude $A(x_s)$ fitted to the distribution
shown in (a), (c) distribution of $x_s$ giving the highest value
of $A(x_s)$ when the experiment was repeated 1000 times. 
\label{amf11y1n}}
\end{center}
\end{figure}

Deteriorating the proper time resolution form the 'nominal' 0.07~ps
to 0.11~ps reduces the $\xsm$ to 27, as could be expected from formula
\ref{basic}. The corresponding summary plot is shown in figure \ref{amf11y1n}.

\begin{figure}
\begin{center}
\mbox{\epsfig{file=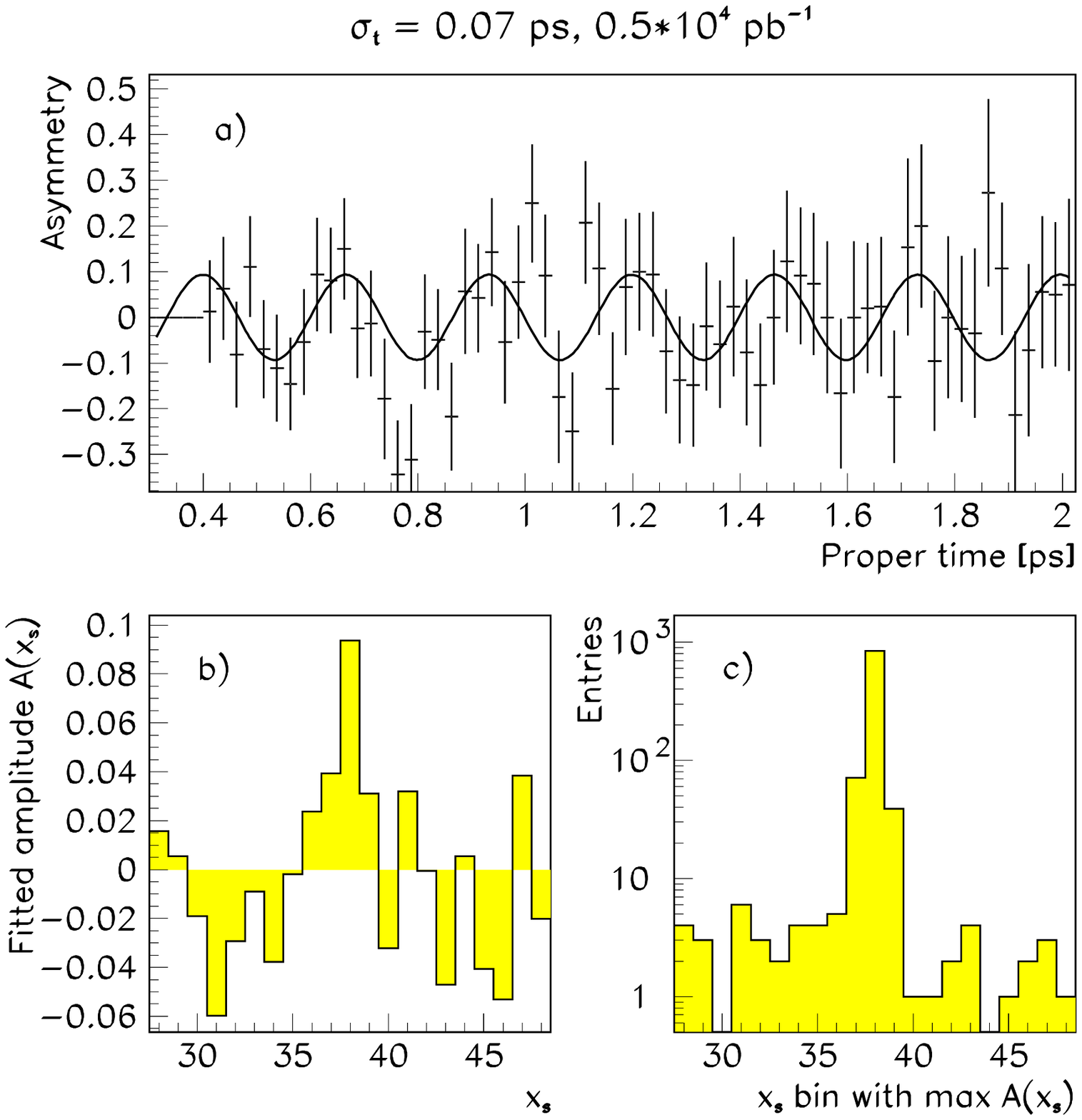,width=17.0cm}}
\caption{$x_s = 38$, $\sigma_t = 0.07$~ps 
and $\lint = 0.5 \cdot \lintu$; (a) time-dependent asymmetry distribution 
(single experiment), (b) amplitude $A(x_s)$ fitted to the distribution
shown in (a), (c) distribution of $x_s$ giving the highest value
of $A(x_s)$ when the experiment was repeated 1000 times. 
\label{amf07y0.5n}}
\end{center}
\end{figure}

If $\st$ has the nominal value and the integrated luminosity is reduced
to $0.5 \cdot \lintu$ the $\xsm$ is less reduced: $\xsm = 38$ (see figure
\ref{amf07y0.5n}). 

\begin{figure}
\begin{center}
\mbox{\epsfig{file=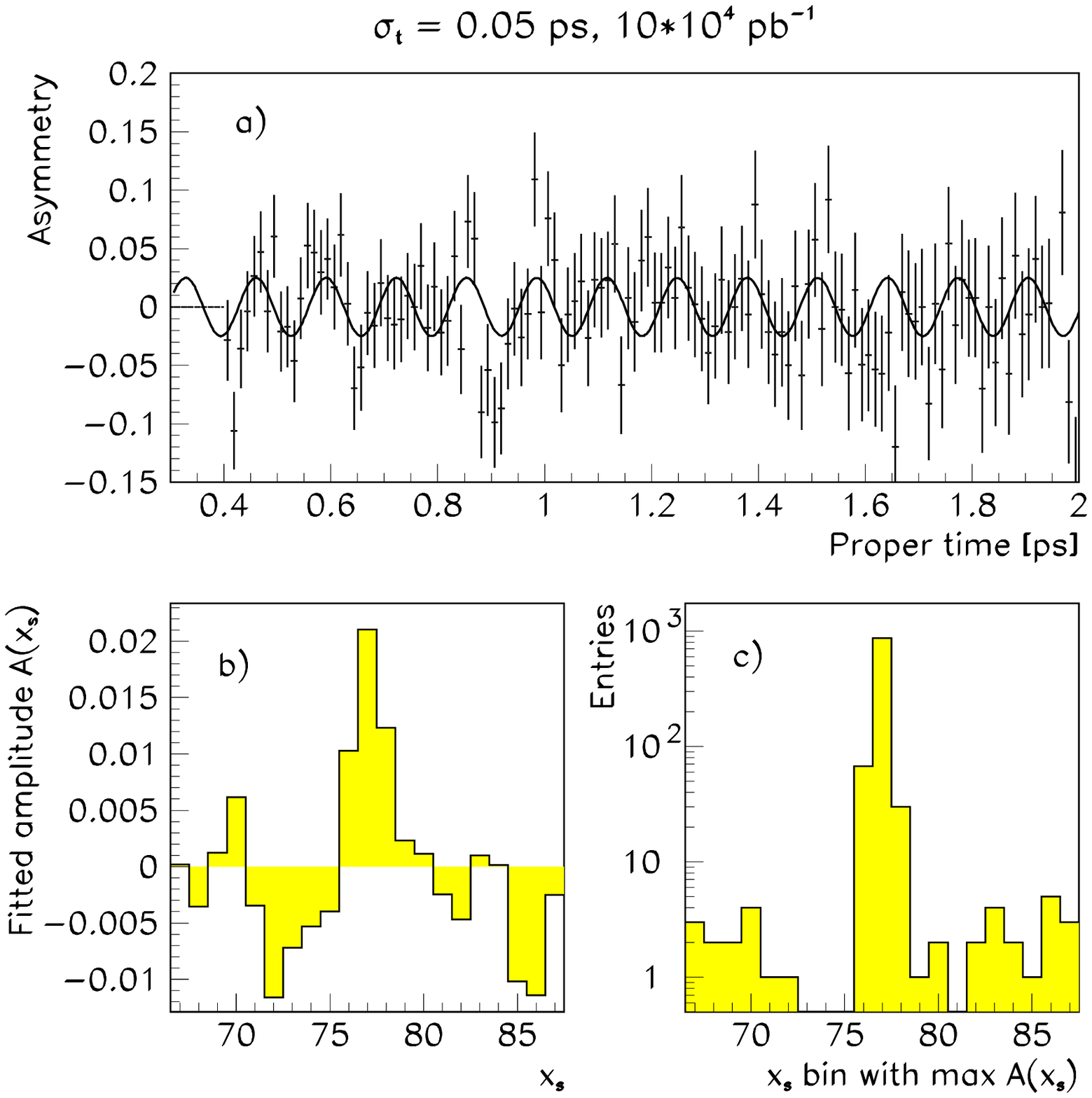,width=17.0cm}}
\caption{$x_s = 77$, $\sigma_t = 0.05$~ps 
and $\lint = 10 \cdot \lintu$; (a) time-dependent asymmetry distribution 
(single experiment), (b) amplitude $A(x_s)$ fitted to the distribution
shown in (a), (c) distribution of $x_s$ giving the highest value
of $A(x_s)$ when the experiment was repeated 1000 times. 
\label{amf05y10n}}
\end{center}
\end{figure}

A~full scan of the two parameters also gives a~set of values: 
$\st = 0.05$~ps and $\lint = 10^5 {\mathrm pb}^{-1}$.
Such a~combination gives $\xsm = 77$ (Fig.~\ref{amf05y10n}). 

\begin{figure}
\begin{center}
\mbox{\epsfig{file=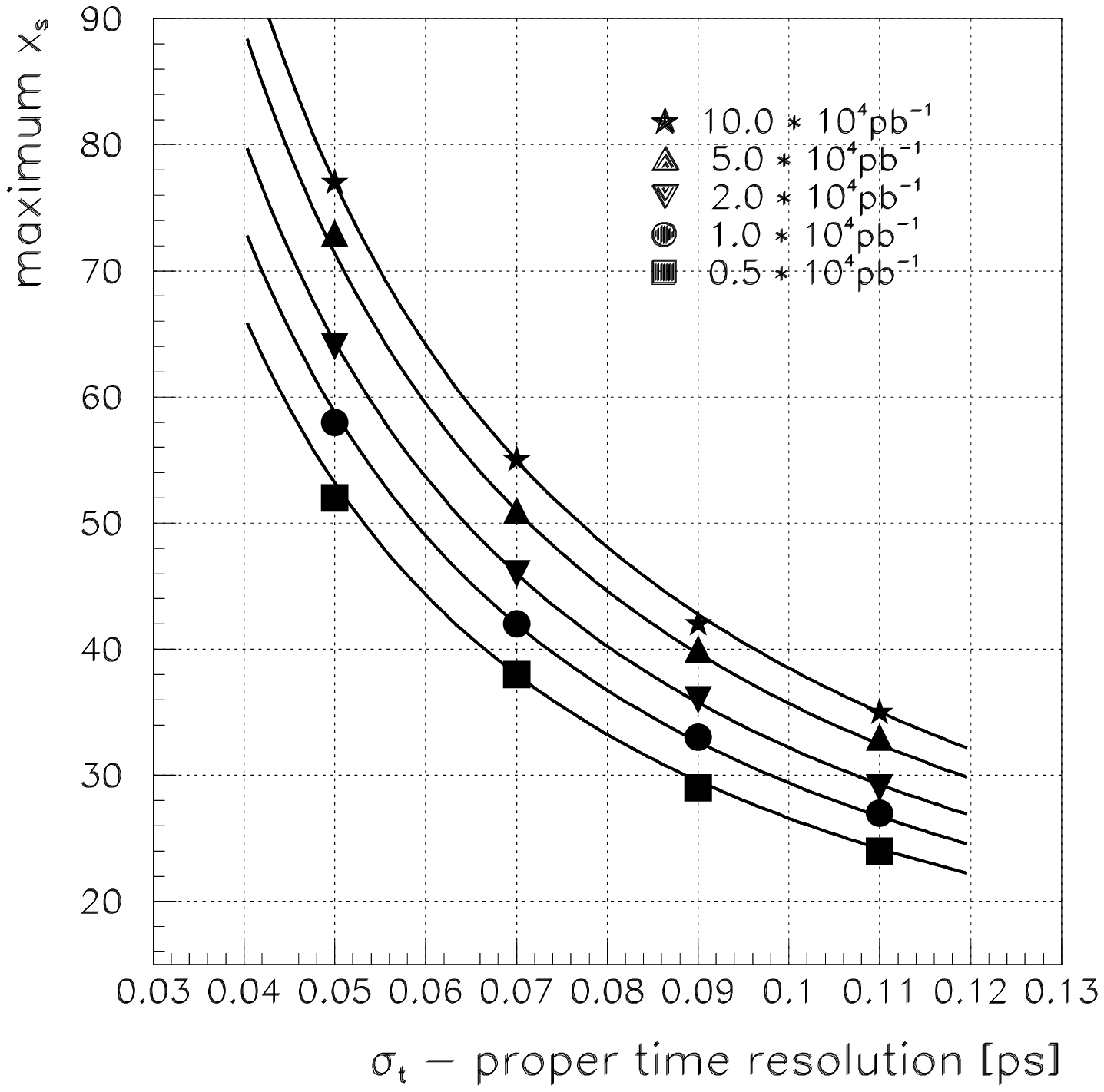,width=15.0cm}}
\caption{Sensitivity range of the \mbox{ATLAS} experiment for the 
$x_s$ measurement, as a~function of the proper-time resolution
$\sigma_t$, for various integrated luminosities. The points 
come from Monte-Carlo calculations, the solid lines show 
$1/\sigma_t$ dependence. 
\label{xsmnt}}
\end{center}
\end{figure}

A~summary plot showing the $\xsm$ versus $\st$ for various integrated luminosities
is shown in figure \ref{xsmnt}. The expected dependence given by the formula 
\ref{basic}, shown with solid lines in the figure, is well confirmed.

\begin{figure}
\begin{center}
\mbox{\epsfig{file=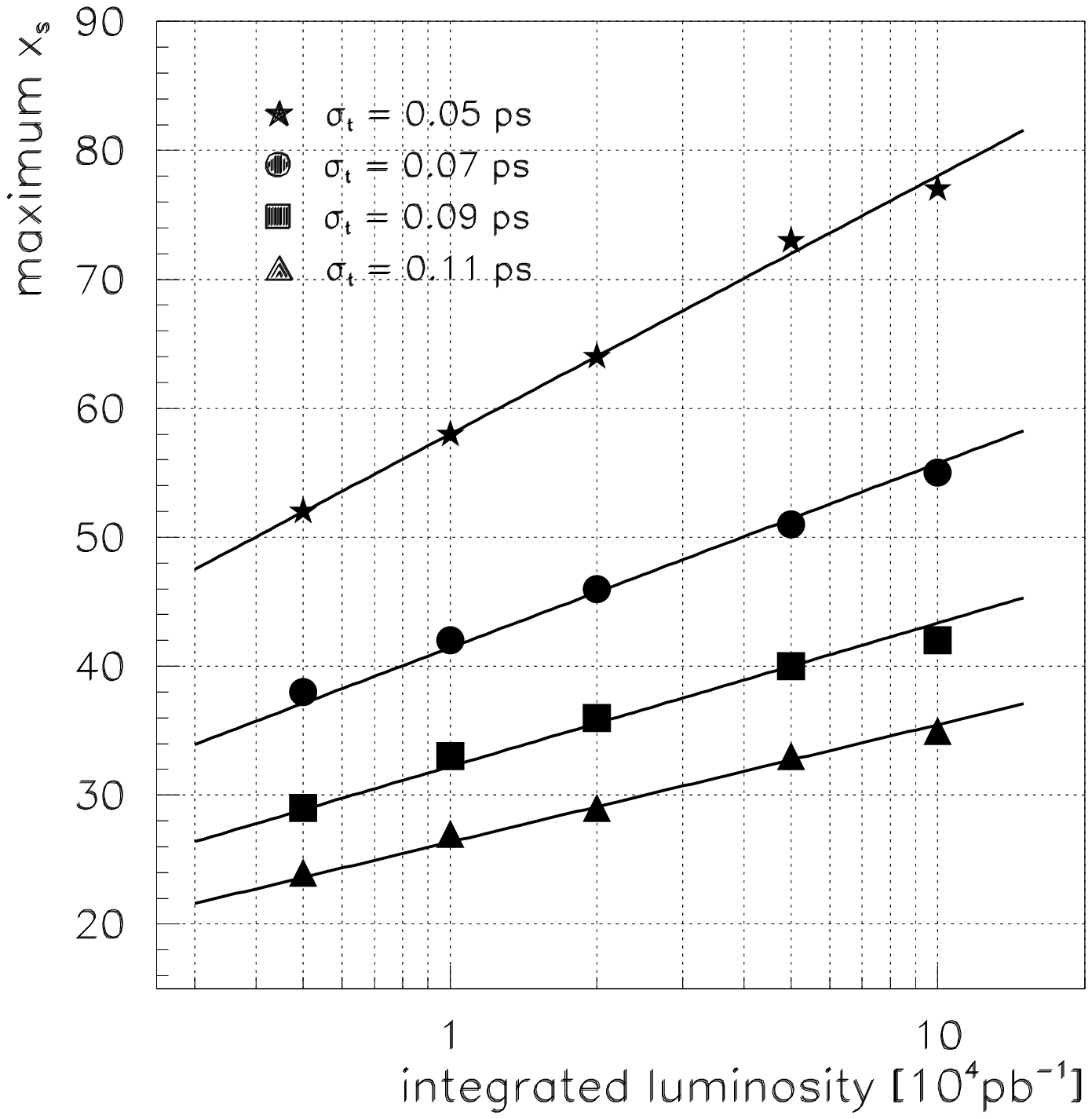,width=15.0cm}}
\caption{Sensitivity range of the \mbox{ATLAS} experiment for the 
$x_s$ measurement, as a~function of the integrated luminosity,
for various proper-time resolutions $\sigma_t$. The points 
come from Monte-Carlo calculations, the solid lines show 
the $x_s^{max} \simeq (2.90/\sigma_t) \cdot (1+0.15\ln(\lint))$ 
parametrisation, where $\sigma_t$ is in ps and \lint is
in units of \lintu. 
\label{xsmnl}}
\end{center}
\end{figure}

The same summary in another projection, $\xsm$ versus \lint for different 
values of \st, is shown in figure \ref{xsmnl}. It is found, 
that the \mbox{ATLAS}
range for the \xs measurement can be parameterized in the following
way:
\begin{equation}
\xsm \simeq \frac{2.90}{\sigma_t}
\left[{1+0.15\ln\left({\lint} \right) }\right]
\label{parameterization}
\end{equation}
where $\st$ is given in ps and \lint in units of \lintu. The solid lines in 
figure \ref{xsmnl} show the prediction of the above formula. 

\section{Dependence on signal-to-background ratio \label{depback}}

\begin{figure}
\begin{center}
\mbox{\epsfig{file=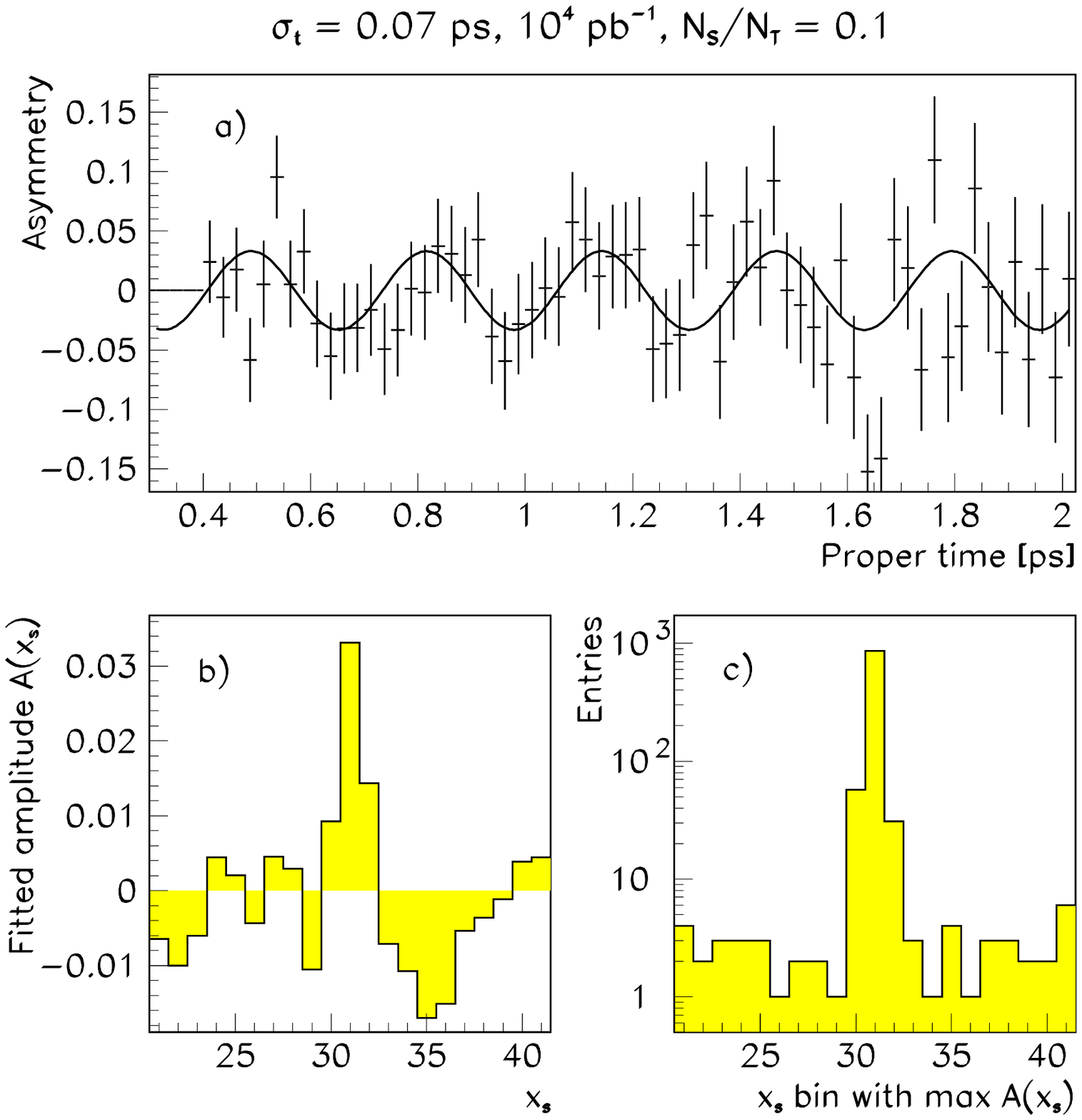,width=17.0cm}}
\caption{$x_s = 31$, $\sigma_t = 0.07$~ps 
and $\lint = \lintu$, $N_B/N_S = 1.0$; 
(a) time-dependent asymmetry distribution 
(single experiment), (b) amplitude $A(x_s)$ fitted to the distribution
shown in (a), (c) distribution of $x_s$ giving the highest value
of $A(x_s)$ when the experiment was repeated 1000 times. 
\label{amf07b9n}}
\end{center}
\end{figure}

The scan of parameters $\st$ and \lint described in the previous section
was done for the background-to-signal ratio of 1. By increasing 
the number of background events such that $N_S/(N_B + N_S)$ becomes equal 0.1 
we reduce $\xsm$ to 31 for the 'nominal' parameters
$\st = 0.07$~ps and $\lint = \lintu$ (see figure \ref{amf07b9n}). 

\begin{figure}
\begin{center}
\mbox{\epsfig{file=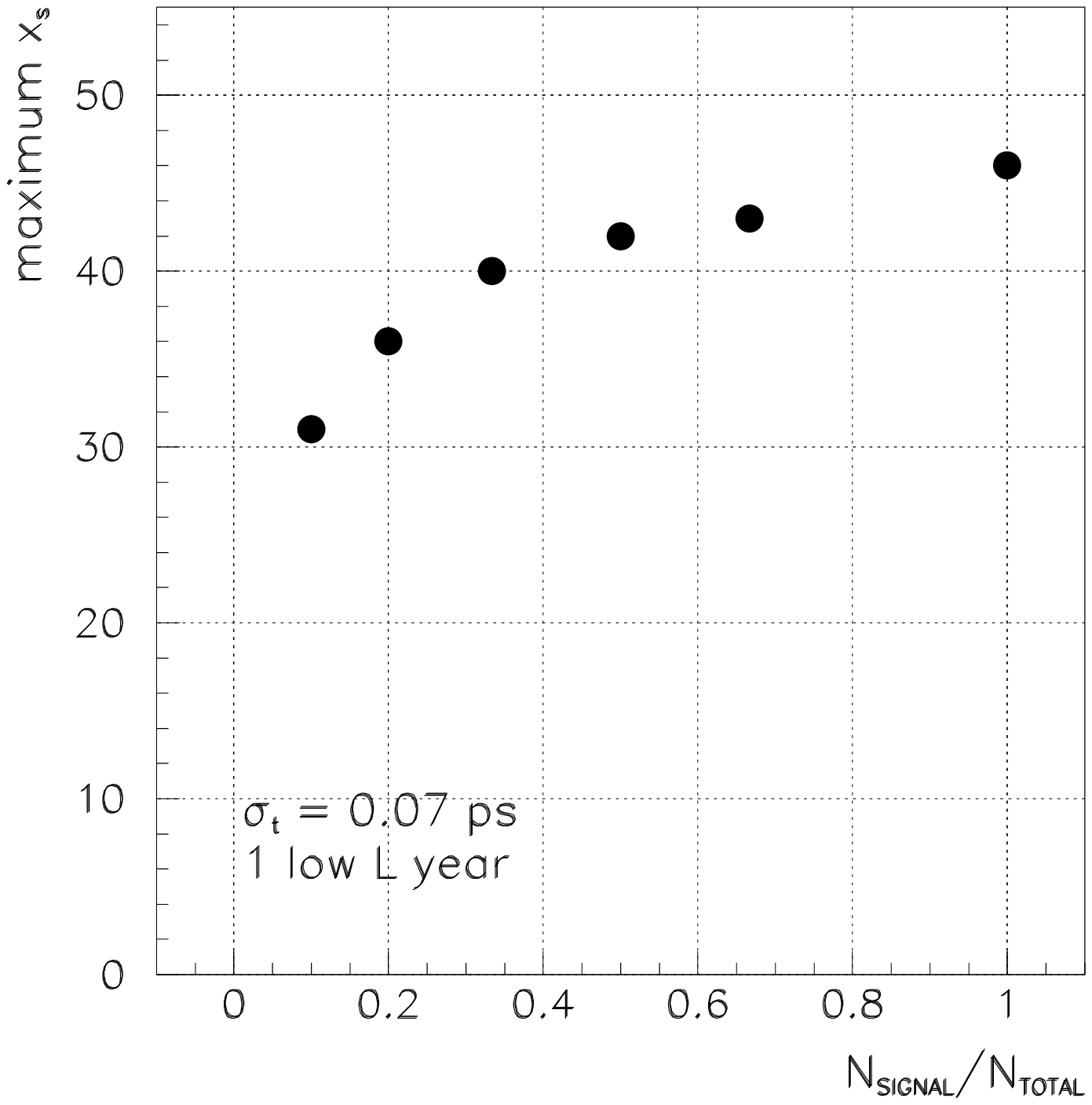,width=12.0cm}}
\caption{Sensitivity range of the \mbox{ATLAS} experiment for the 
$x_s$ measurement, as a~function of the signal content of the sample,
for $\sigma_t = 0.07$~ps and $\lint = \lintu$.}
\label{xsmnb}
\end{center}
\end{figure}

A summary plot showing the dependence of $\xsm$ on background 
is shown in figure \ref{xsmnb}. $N_S$ was kept constant at its predicted 
value for $\lint = \lintu$ ($N_S = 4480$, see section \ref{params}), 
while the number of background event was varied between $N_B = 0$ and 
$N_B = 9 \times N_S$. It can be observed, that the dependence
of $\xsm$ on background is not strong. It should be noted however, that 
background had no asymmetry in the simulation and a~simple exponential 
dependence $d N_B/dt \sim \exp(-t/\tau)$ was assumed.  

\clearpage
\section{Conclusions}

It is justified theoretically, after \cite{hgm}, that the sensitivity 
range $\xsm$ is proportional to $1/\st$ (where $\st$ is the resolution in
the B-decay proper time), for any given integrated luminosity and 
signal-to-background ratio, independently of the method of defining 
the sensitivity range. 

With the methods described in section \ref{sectrange} it is found that 
$\xsm = 42$ for nominal \mbox{ATLAS} parameters. The Monte-Carlo study 
done for various integrated luminosities and for different $\st$ values 
shows, that \mbox{ATLAS} sensitivity range can be parameterized as 
$x_s^{max} \simeq (2.90/\sigma_t) \cdot (1+0.15\ln(\lint))$, where
$\st$ is in ps and $\lint$ is in units of $\lintu$. 
It is also found, that the dependence of $\xsm$ on background is 
relatively weak.


\flushleft

\end{document}